# Lost in Data: How Older Adults Perceive and Navigate Health Data Representations

Peterson Jean[1], Emma Murphy1, and Enda Bates[2]

[1]School of Computer Science, Technological University Dublin
jean.peterson@tudublin.ie
[2]School of Electronic and Electrical Engineering, Trinity College Dublin

**Abstract.** As the ageing population grows, older adults increasingly rely on wearable devices to monitor chronic conditions. However, conventional health data representations (HDRs) often present accessibility challenges, particularly for critical health parameters like blood pressure and sleep data. This study explores how older adults interact with these representations, identifying key barriers such as semantic inconsistency and difficulties in understanding. While research has primarily focused on data collection, less attention has been given to how information is output and understood by end-users. To address this, an end-user evaluation was conducted with 16 older adults (65+) in a structured workshop, using think-aloud protocols and participatory design activities. The findings highlight the importance of affordance and familiarity in improving accessibility, emphasising the familiarity and potential of multimodal cues. This study bridges the gap between domain experts and end-users, providing a replicable methodological approach for designing intuitive, multisensory HDRs that better align with older adults' needs and abilities.

**Keywords**: Health data representation · Accessibility · Older adults · End-user evaluation

## 1 Introduction

The global population is ageing rapidly, leading to an increased prevalence of chronic health conditions. While older adults aim for better independence, a primary challenge remains the effective self-management of these conditions. Chronic illnesses often increase the risk of emergency situations [3]. Many key health parameters, like blood pressure , are particularly critical due to their reactivity and association with multiple underlying conditions. their variability requires continuous observation over time, as delayed intervention can have severe consequences. Technology plays an essential role in tracking such physiological data, yet its effectiveness depends not only on data collection but also on how this information is conveyed to end-users. The adoption of digital health technologies among older adults has grown significantly. According to the OECD [5], there has been a notable increase in technology use, but such adoption is influenced by many factors such as education, income, and, most significantly, age. Additionally, older adults form a highly heterogeneous group with diverse cognitive, physical, and technological capabilities, further complicating universal adoption [5, 4].

Wearable devices have emerged as a supportive solution for health monitoring, with extensive research efforts focused on improving data collection and sensor accuracy. However, significantly less attention has been given to how this data is presented to users. The way information is structured directly impacts comprehension, and poor representation may limit usability, particularly for older adults. While accessibility guidelines exist for accessible interfaces, there is limited research on guidelines application to health data representation for this older demographic [4]. Moreover, much of the work on alternative health data representation often targets experts rather than end-users like older adults. Our initial expert reviews highlighted critical accessibility issues in conventional health data representation. However, these insights alone are insufficient, as they do not capture the lived experiences of older adults. A major gap remains in understanding their specific challenges when interpreting critical physiological health parameters and how these representations fit into their daily lives. Effective representation must go beyond numerical values to provide context and meaning, ensuring that older adults can not only read the data but also understand its relevance. Addressing these challenges requires a user-centred approach that prioritises accessibility, clarity, and applicability, emphasising not only domain expertise in accessibility but also the needs and experiences of older adults themselves.

## 2  Methods

This study evaluated end-user experiences with the four most common conventional health data representations (HDRs) as a scoped baseline previously assessed by domain experts. The focus was on accessibility, ease of use, and comprehension, particularly for older adults. Two screens displayed blood pressure data, while two represented sleep patterns. With ethics approval obtained, 16 older adults (M=3, F=13) aged 65+ were recruited through a local voluntary service organisation in Dublin. A two-hour workshop was conducted at TU Dublin, where participants were divided into three groups (approximately six per group. The workshop was facilitated by a lead facilitator and three supporting facilitators. HDRs were presented as mobile app screens, displayed on tablets (one iPad and three Samsung S8 Tabs) and printed on paper at actual screen size for accessibility. The workshop included three activities: (1) a think-aloud evaluation using scenarios and persona archetypes, (2) a brainstorming session using the "I Like, I Wish, What If" framework to elicit feedback and design suggestions, and (3) the collection of participant demographic data and health parameter usage. All sessions were recorded, anonymised, transcribed, and analysed. The transcripts were analysed using Clarke and Braun's [2] thematic analysis to identify themes.

## 3  Results

From the demographic survey and health data usage results, participants ranged in age from 65 to 103 years. Most older adults (14 out of 16) own a smartphone. The most common activities included texting, making phone calls, conducting internet searches, and using apps like WhatsApp. Half of the participants (8 out of 16) monitor some form of health parameter, with steps being the most frequently tracked, using various downloaded health apps. Seven out of 16 participants use health apps, with Fitbit and pedometers being the most commonly mentioned. This aligns with the research findings reported by the OECD regarding the higher adoption of technology and the downloading of health-related apps, even though older adults are not necessarily actively utilising them [3]. This supports the identification of a gap in self-management support in Ireland and highlights its emergence as a global issue. Considering age-related conditions, there was no clear correlation between age and the likelihood of measuring specific health parameters; both younger (71) and older (95+) participants track some form of health parameters. This also demonstrates the heterogeneous characteristics of the demographic. From the thematic analysis, insights from the codes reveal numerous similarities with findings from domain experts regarding accessibility issues. Patterns such as confusion, a lack of information, and indications of insufficient digital health literacy were consistently observed. For instance, a participant expressed sentiments such as, "That doesn't tell you anything. Yeah, don't tell anything so,..." or, "[I have a] Fitbit, and I would look at that, and it tells me that my heart is 55, resting is 53. That's enough." Here, we could argue that this highlights a sense of familiarisation as recollection, reducing cognitive load when interpreting the information. These statements indicate that while older adults may not universally be affected by the initial layer of the digital divide related to accessing technology, challenges in digital health literacy and understanding data representation continue to impair their effective use of such essential tools. When asked, older adults expressed familiarity and awareness of existing alternative data representations using audio, such as alarms and haptics like phone vibrations.

## 4      Discussion

There is a strong emphasis on wearable data collection, yet significantly less attention is given to how this data is presented to users. A key challenge for older adults when interacting with health data representations (HDRs) is the inconsistency of semantic elements, leading to confusion. Participants demonstrated an awareness of alternative multimodal ways of presenting data, such as vibrations or alarms, supporting the argument that affordance plays a crucial role in HDR interaction. Cajamarca et al. [1] similarly suggest that older adults prefer familiarity to simplify interactions. Wallace et al. [6] further argue that affordances in objects or artefacts inherently convey meaning, acting as a conduit for information across time, space, or memory. These findings emphasise the need for intuitive and accessible health data representations that align with older adults' cognitive abilities and learning styles. This study suggests that future HDR designs should go beyond predominantly visual formats by integrating sensory adaptations that leverage familiarity and affordance-based cues.

## 5      Conclusion/Implications for the AT Field

This study highlighted the accessibility challenges older adults encounter with conventional HDRs, particularly regarding blood pressure and sleep data. It bridged the gap between domain experts in the AT field and end-users, offering an inclusive methodological approach to address age-related barriers of HDRs. Beyond contributing to multisensory prototype requirements, it provides a replicable approach for exploring other health data representations for various health parameters.

**Acknowledgments.**

This work was conducted with the financial support of the Research Ireland Centre for Research Training in Digitally-Enhanced Reality (d-real) under Grant No. 18/CRT/6224.